\documentclass[showpacs,prd,amsfonts]{revtex4}

\usepackage[dvips]{graphicx}
\def\simless{\mathbin{\lower 3pt\hbox
{$\rlap{\raise 5pt\hbox{$\char'074$}}\mathchar"7218$}}}   
\def\simmore{\mathbin{\lower 3pt\hbox
{$\rlap{\raise 5pt\hbox{$\char'076$}}\mathchar"7218$}}}   
\newcommand{\be}{\begin{equation}}
\newcommand{\ee}{\end{equation}}

\begin{document}

\title{Spherically symmetric, static spacetimes in a tensor-vector-scalar theory}
\author{Dimitrios Giannios \email{giannios@mpa-garching.mpg.de}
}

\affiliation{Max Planck Institute for Astrophysics, Box 1317, D-85741 Garching, 
Germany
}

\date{\today}

\begin{abstract}
Recently, a relativistic gravitation theory has been proposed  [J. D. Bekenstein, Phys. Rev.
D {\bf 70}, 083509 (2004)] that gives the Modified Newtonian Dynamics (or MOND) in the weak 
acceleration regime. The theory is based on three dynamic gravitational fields and succeeds in 
explaining a large part of extragalactic and gravitational lensing phenomenology without invoking 
dark matter. In this work we consider the strong gravity regime of TeVeS. We study spherically 
symmetric, static and vacuum spacetimes 
relevant for a non-rotating black hole or the exterior of a star. Two branches of solutions
are identified: in the first the vector field is aligned with the time direction while in the second
 the vector field has a non-vanishing radial component. We show that 
in the first branch of solutions the $\beta$ and $\gamma$ PPN coefficients in TeVeS are identical 
to these of general relativity
(GR) while in the second the $\beta$ PPN coefficient differs from unity violating
observational determinations of it (for the choice of the free function $F$ of the theory made in 
Bekenstein's paper). For the first branch of solutions, we derive analytic expressions for 
the physical metric and discuss their implications. Applying these solutions to the case of black holes, 
it is shown that they violate causality (since they allow for superluminal propagation of metric, 
vector and scalar waves) in the vicinity of the event horizon and/or that they are characterized by  
negative energy density carried by the fields.
 
\end{abstract}


\maketitle

\section{Introduction} \label{intro}

On cosmological scales, Newtonian gravitational theory under-predicts
the acceleration of stars and gas. Furthermore, galaxies and clusters of galaxies show
 anomalously large gravitational lensing when only baryonic matter is taken into account.
A natural ``cure'' for these discrepancies is to assume the existence
of dark matter, which dominates over visible
matter~\cite{Zwicky, Rubin, Ostriker, Blumenthal, Trimble}. Such dark matter 
might solve the ``missing mass'' problem within
the standard theory of gravity (i.e., general relativity or GR). For this picture to be complete,
however, the origin of dark matter needs to be identified.   

A second approach to the acceleration discrepancy and lensing anomaly
is to look for alternative theories of gravity, that modify GR on large scales.
Among many other attempts, the MOND paradigm has been 
proposed~\cite{Milgrom1, Milgrom2}.
It is characterized by an acceleration scale $a_0$ so that
\be
\tilde{\mu}(|\vec{a}|/a_0)\vec{a}=-\nabla \Phi_{\rm N},
\ee  
where $\Phi_{\rm N}$ is to be understood as the Newtonian potential,
$\tilde{\mu}(x)=x$ for $x\ll 1$, while  $\tilde{\mu}(x)=1$ for $x\gg 1$.

This empirical law has been very successful in explaining
the rotation curves in a large number of (spiral, low surface brightness
and elliptical) galaxies using the observed distributions of gas and
stars as input \cite{Kent,Milgrom3,Begeman}. MOND can also 
explain  the observed correlation between
the infrared luminosity of a disk galaxy $L_K$ and the asymptotic rotational
velocity $v_{a}$ (i.e. the the Tully-Fisher law: $L_K\propto v_a^4$~\cite{Tully}), 
on the (suggested by population synthesis models) assumption that 
the $M/L_K$ ratio is constant.  

However, MOND is merely a prescription for gravity and not
a self consistent theory. It violates, for example,  conservation of momentum and 
angular momentum  and does not provide the formulation to describe
light deflection or to build a cosmological model. A  
theory of gravity is needed that has the MOND characteristics 
in the weak acceleration limit but also has full predictive power.
  
In this context, a new relativistic theory of gravity has been introduced
by Bekenstein~\cite{Bek}.
It consists of three dynamical gravitation fields: a tensor field
($g_{\alpha \beta}$), a vector field ($U_{\alpha}$) and a scalar
field ($\phi$) leading to the acronym TeVeS.
The theory involves a free function $F$, a length scale $\ell$ (that can
be related to $a_0$) and two positive dimensionless constants $\kappa$, K. 

TeVeS gives MOND in the weak acceleration limit (and therefore 
inherits the successes of MOND on large scale), makes similar
prediction on gravitational lensing as GR (with dark matter) and provides
a formulation for constructing cosmological models.      
 As a drawback however, one can mention that TeVeS  still appears to need 
dark matter to address the cosmological matter problem, i.e., the fact that
observations require that the source term of Friedmann's equations is a 
factor of $\sim 6$ the baryonic matter density. 

A self consistent theory must be causal (i.e. not to allow
for superluminal propagation of any measurable field or energy) 
and in TeVeS this is the case provided that
$\phi>0$ (~\cite{Bek}; Section VIII). It can be shown
that, for a range of initial conditions, FRW cosmological
models with flat spaces in TeVeS expand for ever with
$0<\phi\ll 1$ throughout. Moreover, in the vicinity of a 
star embedded in this cosmological background $\phi$ is still
positive. So, in a wide range of environments, TeVeS has been shown
to be causal. 

The predictions of the theory have thus been explored to some extent 
for a range of strengths of the gravitational field: from the
MOND limit to the Post-Newtonian corrections in the inner solar system.
The strong gravity regime (e.g. in the vicinity of a black hole
or a neutron star) of TeVeS has not been studied. This regime is a topic of this work.
The motivation for this work is two-fold. First, it is interesting
to see how the physics of compact objects (i.e. black holes, neutron stars)
differ in TeVeS with respect to GR and what constraints (if any) observations can
put on its free parameters. Second, one can check the
consistency of the theory (e.g., its causality, positivity of energy carried by the fields) in 
these extreme conditions.    

In Section II, we summarize the fundamentals of TeVeS and in Section III
we consider its strong gravity limit. We limit ourselves to static, spherically symmetric 
and vacuum spacetimes relevant for a non-rotating black hole or the exterior of
a star.  Two branches of solutions are identified: in the first the vector 
field is aligned with the time direction while in the (not previously explored) second branch
the vector field has a non-vanishing radial component.  We show that the $\beta$ and $\gamma$ 
PPN coefficients in TeVeS are identical to these of GR in the first branch of solutions while the 
$\beta$ PPN
coefficient differs in the second. For the choice of the free 
function $F$ made in Ref.~\cite{Bek}, we find that TeVeS predicts a value for $\beta$ that is in 
conflict with recent 
observational determinations of it. In Section IV, we consider the first branch of solutions and derive 
exact solutions for the metric for arbitrary values of the parameters of the theory. The observational
properties of the black holes in TeVeS are discussed in Sec.~V along with
the issue of superluminal propagation of waves in the black hole vicinity. Conclusions are
given in Sec.~VI.  

\section{The basic equations of TeVeS}

TeVeS is based on three dynamical gravitational fields: a tensor
field (the Einstein metric $g_{\alpha \beta}$), a 4-vector field
$U_{\alpha}$ and a scalar field $\phi$ with an additional
nondynamical scalar field $\sigma$. The physical metric $\tilde{g}_{\alpha 
\beta}$ in TeVeS is connected to these fields through the expression 
\be  
\tilde{g}_{\alpha \beta}\equiv e^{-2\phi}g_{\alpha \beta}-2U_{\alpha}U_{\beta}
\sinh (2\phi).
\label{physmetric}
\ee 

The total action in TeVeS is the sum of 4 terms $S_g$, $S_s$, $S_U$
and $S_m$ (see Ref.~\cite{Bek}), where $S_g$ is identical to the Hilbert-Einstein 
action and is the part that corresponds to the tensor field,
while $S_s$, $S_U$, $S_m$ are the actions of the two scalar fields,  
 the vector field and the matter respectively.   
The basic equations of TeVeS are derived by varying the total action
S with respect to $g^{\alpha \beta}$, $\phi$, $\sigma$, $U_{\alpha}$.

Doing so for $g^{\alpha \beta}$, one arrives to the metric equations 
\be
G_{\alpha \beta}=8\pi G[\tilde{T}_{\alpha \beta}+(1-e^{4\phi})U^{\mu}
\tilde{T}_{\mu(\alpha} U_{\beta)}+\tau_{\alpha \beta}]+\Theta_{\alpha \beta},
\label{metrice}
\ee
where a pair of indices surrounded by parenthesis stands for symmetrization,
 i.e. $A_{(\alpha}B_{\beta)}=A_{\alpha}B_{\beta}+A_{\beta}B_{\alpha}$,
the $G_{\alpha \beta}$ denotes the Einstein tensor for $g_{\alpha \beta}$,
 $\tilde{T}_{\alpha\beta}$ is the energy momentum tensor
and
\begin{eqnarray}
\tau_{\alpha\beta} \equiv
\sigma^2\Big[\phi_{,\alpha}\phi_{,\beta}-{\scriptstyle 1\over
\scriptstyle 2}g^{\mu\nu}\phi_{,\mu}\phi_{,\nu}\,g_{\alpha\beta}-
U^\mu\phi_{,\mu}\big(U_{(\alpha}\phi_{,\beta)}-
{\scriptstyle 1\over \scriptstyle
2}U^\nu\phi_{,\nu}\,g_{\alpha\beta}\big)\Big]
\label{tau}
\nonumber \\ 
-{\scriptstyle 1\over\scriptstyle 4}G
\ell^{-2}\sigma^4 F(kG\sigma^2)  g_{\alpha\beta},
\end{eqnarray}
\be
\Theta_{\alpha\beta}\equiv
K\Big(g^{\mu\nu} U_{[\mu,\alpha]}
U_{[\nu,\beta]} -{\scriptstyle 1\over \scriptstyle 4}
g^{\sigma\tau}g^{\mu\nu}U_{[\sigma,\mu]}
U_{[\tau,\nu]}\,g_{\alpha\beta}\Big)-
\lambda U_\alpha U_\beta,
\label{Theta}
\ee 
where a pair of indices surrounded by brackets stands for antisymmetrization,
 i.e. $A_{[\alpha}B_{\beta]}=A_{\alpha}B_{\beta}-A_{\beta}B_{\alpha}$.

Similarly one derives a scalar equation that can be brought into the form
\be
\left[\mu \left(k\ell^2 h^{\mu\nu}\phi_{,\mu}\phi_{,\nu}\right)
h^{\alpha\beta}\phi_{,\alpha} \right]_{;\beta}= 
kG\big[g^{\alpha\beta} + (1+e^{-4\phi}) U^\alpha
U^\beta\big] \tilde T_{\alpha\beta},
\label{s_equation}
\ee
where $h^{\alpha\beta}\equiv g^{\alpha\beta}-U^{\alpha}U^{\beta}$ and
 $\mu(y)$ is defined by
\be
 -\mu F(\mu ) -{\frac{1}{2}}\,
\mu ^2\dot{F}(\mu ) = y,
\label{F}
\ee 
where $\dot{F}\equiv dF/d\mu$.
The scalar field $\sigma$ is given by
\be 
kG\sigma^2=\mu (k\ell^2h^{\alpha\beta}\phi_{,\alpha}\phi_{,\beta}).
\label{sigma}
\ee  
Note that the form of the function $F(\mu)$ [or equivalently $y(\mu)$]
is not predicted by the theory and is essentially a free function.
In next Section  we give the form used in Bekenstein's
 paper. As it will turn out, however, the
results derived here are essentially 
independent of the exact choice of $F(\mu)$ and
quite general. On the other hand, our final conclusions do depend on the
choice of $F$, since it influences the way in which observations put
 constraints on the parameters of the theory (see for example Sec.~II~A).  

Finally, the vector equation is derived through
variation of $S$ with respect to $U_\alpha$ 
\be K
U^{[\alpha;\beta]}{}_{;\beta}+\lambda
U^\alpha+8\pi G\sigma^2
U^\beta\phi_{,\beta}g^{\alpha\gamma}\phi_{,\gamma} =
8\pi G (1-e^{-4\phi}) g^{\alpha\mu} U^\beta 
\tilde T_{\mu\beta},
\label{vectoreq}
\ee where $U^\alpha\equiv g^{\alpha\beta}U_{\beta}$ and $\lambda$ is 
a Lagrange multiplier.   These 4 equations determine $\lambda$ and
 three of the components of $U^\alpha$ with the fourth being determined 
by the normalization of the vector field 
\be
g^{\alpha \beta}U_{\alpha}U_{\beta}=-1.
\label{norm}
\ee

\subsection{The function F}

The function $F(\mu)$ (or equivalently of $y(\mu)$)
is a free function  since there is no theory for
it. One has large freedom in choosing  the form of $F$, each to be checked on 
 implications for cosmological models, galactic rotation
curves and constraints from  measurements in the outer solar system. 
Bekenstein in Ref.~\cite{Bek} made the following choice
\be
y(\mu)=\frac{3}{4}\frac{\mu^2(\mu-2)^2}{1-\mu}
\label{ymu}
\ee
which, using Eq.~(\ref{F}) leads to 
\be
F(\mu)=\frac{3}{8}\frac{\mu(4+2\mu-4\mu^2+\mu^3)+4\ln (1-\mu)}{\mu^2}.
\label{fmu}
\ee 
It can be shown that the range $0<\mu<1$ (i.e., $y>0$) is relevant for
quasistationary systems and $2<\mu<\infty$ (i.e., $y<0$) 
for cosmology. For this specific choice of $F(\mu)$, one can put
a lower limit on the value of the $\kappa$ parameter of the theory
so that it is not in conflict with the measured motions of planets of the
outer solar system (see Ref.~\cite{Bek}; Sec. IV). On the other hand, small
values of $\kappa$ are relevant for cosmological models.
Together, these constraints indicate a value of $\kappa$ around $\sim 0.03$. It should be stressed
that it depends on the specific choice of the form of
$F(\mu)$.

 The Newtonian limit of a spherically symmetric system has been explored in  
Sec. IV C of Ref.~\cite{Bek}, where it is shown that for gravitational accelerations
$|\vec{a}|/a_0\gg 8\pi^2/\kappa^2$  the quantity $y\to \infty$ and 
consequently $\mu\to 1$. As an arithmetic example, at earth's and Mercury's orbit 
$\mu$ differs from unity by about $2\cdot 10^{-6}$ and $5\cdot 10^{-8}$ respectively
for the specific choice (\ref{fmu}) of function $F$ and $\kappa=0.03$.    
Since in this study we focus on the strong gravity limit, we can
safely take $\mu=1$ and, therefore, [see Eq.~(\ref{sigma})] 
\be
\sigma^2=\frac{1}{\kappa G}.
\label{sigma2}
\ee
Strictly speaking, $\mu$ has been shown to be of order unity in the Newtonian limit 
but not necessary in the relativistic limit. However, using the analytic solutions derived in Sec.~IV,
we have checked that taking $\mu=1$ is an excellent approximation also in this limit.

\section{Spherical symmetric, static spacetimes  in TeVeS}

From this point on we focus on the strong gravity limit of TeVeS
and explore the spacetime in the vicinity of a  spherically
symmetric mass.
The isotropic form of a spherical symmetric, static metric is
\be
g_{\alpha \beta}dx^\alpha dx^\beta=-e^\nu dt^2+e^\zeta (dr^2+r^2d\theta^2
+r^2\sin^2\theta d\varphi^2), 
\label{metric} 
\ee
where both $\nu$, $\zeta$ are only functions of $r$.  For the  static 
system of our case the vector field has two non-vanishing components
\be
U^\alpha=(U^t,U^r,0,0),
\label{Ufield}
\ee     
where $U^t$ and $U^r$ are functions of the radial coordinate. 

Taking $\phi=\phi(r)$, the scalar-field 
equation~(\ref{s_equation}) in vacuum may be written as 
\be
\frac{e^{-(\nu+3\zeta)/2}}{r^2}[r^2e^{(\nu+3\zeta)/2}\phi'(e^{-\zeta}-(U^r)^2)]'=0,
\label{phi_s}
\ee
where a prime stands for ordinary derivative with respect to $r$.

Equation~(\ref{phi_s}) can be integrated once to give 
\be
\phi'=\rm C \frac{e^{-(\nu+3\zeta)/2}}{r^2[e^{-\zeta}-(U^r)^2]},
\label{phiprime}
\ee
where $C$ is an integration constant. 
To determine this constant, one must consider the
source of the gravitational field. A 
analysis relevant for an extended source (e.g. a star) is given in
Bekenstein~\cite{Bek} Sec.~V, where it is shown that one can define
the ``scalar'' mass $m_s$ as a (non-negative) particular  integral 
over $\tilde{\rho}$ and $\tilde{P}$ (defined as the proper energy
density and pressure expressed in the physical metric) of the star's matter 
so that the integration constant is given by 
\be
C\equiv \kappa G m_s/(4\pi).
\ee 
   
The $r$ component of the vector equation (\ref{vectoreq}) (the $\theta$ and $\phi$ components 
vanish because of the symmetry of the problem under consideration) can be brought into the form
\be
U^r\Big(\lambda+\frac{\kappa (Gm_s)^2}{2\pi}\frac{e^{-(\nu+4\zeta)}}{r^4[e^{-\zeta}-(U^r)^2]^2}\Big)=0
\label{Ur}.
\ee
This equation shows that there are two cases: either $U^r$ vanishes or one
has a constraint on the Lagrange multiplier $\lambda$. In the former case the 
vector field is aligned with the time direction, while in the latter there is a non-vanishing
radial component of the vector field. Since the mathematical analysis of the two
cases is rather different, we examine them separately.

\subsection{Case I: the vector field is aligned to the time direction}

When $U^r$ vanishes, the vector field is determined by the normalization expression
(\ref{norm}) which yields
\be
U^\alpha=(e^{-\nu/2},0,0,0).
\label{utI}
\ee    
The physical metric is then given by Eq.~(\ref{physmetric}) which reduces to
 $\tilde{g}_{tt}=g_{tt}e^{2\phi}$, $\tilde{g}_{ii}=g_{ii}e^{-2\phi}$. To determine 
$\tilde{g}^{\alpha\beta}$,
 one needs to solve for $\nu$ and $\zeta$ and, through them, for the metric $g_{\alpha\beta}$
 and the scalar $\phi$.   
To this end, the differential equations resulting from the $tt$, $rr$ and $\theta\theta$ 
components of the metric equation (\ref{metrice}) must be solved. This 
is equivalent to the procedure one follows to arrive at the GR solutions.

Since we are looking for vacuum spacetimes, the terms that include the
matter energy density in Eq.~(\ref{metrice}) are zero. So, we are
left with the $\tau_{\alpha\beta}$ and $\Theta_{\alpha\beta}$ terms.   
The $\tau_{\alpha\beta}$  [see Eq.~(\ref{tau})] contains $\phi_{,\alpha}$
terms while the last term depends on the function $F$. In the strong acceleration limit, 
it is possible to show  (see Sec.~V in Ref.~\cite{Bek}) that the last term is completely 
negligible in comparison to the other terms.
This is exactly the limit we are interested in here and so we will
neglect the $F$ term.  
Using Eqs.~(\ref{tau}), (\ref{phiprime})  and (\ref{utI}), we then find
\be
\tau_{tt}=\frac{\kappa Gm_s^2}{32 \pi^2 }\frac{e^{-2\zeta}}{r^4},
\label{tautt}
\ee \be
\tau_{rr}=\frac{\kappa Gm_s^2}{32 \pi^2 }\frac{e^{-(\zeta+\nu)}}{r^4},
\label{trr}
\ee 
\be
\tau_{\theta\theta}=-\frac{\kappa Gm_s^2}{32 \pi^2 }\frac{e^{-(\zeta+\nu)}}{r^2}.
\label{t88}
\ee 

To proceed with the calculation of the $\Theta_{\alpha \beta}$ 
terms [defined by Eq.~(\ref{Theta})], one first needs to compute the Lagrange
multiplier $\lambda_{\rm I}$ (where the index I is used to show that it corresponds 
to the case I), from the $t$ component of Eq.~(\ref{vectoreq})
(the other components vanish because of the symmetry of the problem and because we
study the case where $U^r=0$). Using
Eq.~(\ref{utI}) and that $U^{\beta}\phi_{,\beta}=0$ and 
$\tilde{T}_{\alpha\beta}=0$, we have
\be
\lambda_{\rm I}=-K e^{-\zeta}\Big(\frac{\nu''}{2}+\frac{\nu'\zeta'}{4}+\frac{\nu'}{r}\Big).
\label{lambda}
\ee  
Substituting Eqs.~(\ref{lambda}) and (\ref{Ufield}) in (\ref{Theta}) we get  
\be
\Theta_{tt}=Ke^{\nu-\zeta}\Big(\frac{(\nu')^2}{8}+\frac{\nu''}{2}+\frac{\nu'\zeta'}{4}+
\frac{\nu'}{r}\Big),
\label{thetatt}
\ee \be
\Theta_{rr}=-\frac{K}{8}(\nu')^2,
\label{thetarr} 
\ee
\be
\Theta_{\theta\theta}=\frac{K}{8}(r\nu')^2,
\label{theta88} 
\ee

We can now use the $tt$ and $rr$ components of the metric equation [Eq.~(\ref{metrice})] 
to derive a system of ordinary differential equations for $\zeta$ and $\nu$. Using Eqs.~(\ref{tautt}), 
(\ref{trr}), (\ref{thetatt}) and (\ref{thetarr})
in (\ref{metrice}), and after some rearrangement, one finds that
\be
\zeta''+\frac{(\zeta')^2}{4}+\frac{2\zeta'}{r}=-\frac{\kappa (Gm_s)^2}{4 \pi }\frac{e^{-(\zeta+\nu)}}
{r^4}-K \Big(\frac{(\nu')^2}{8}+\frac{\nu''}{2}+\frac{\nu'\zeta'}{4}+\frac{\nu'}{r}\Big)
\label{eq1}
\ee
and
\be
\frac{(\zeta')^2}{4}+\frac{\zeta'\nu'}{2}+\frac{\zeta'+\nu'}{r}=\frac{\kappa (Gm_s)^2}{4 \pi }
\frac{e^{-(\zeta+\nu)}}{r^4}-K\frac{(\nu')^2}{8}.
\label{eq2}
\ee
These two equations are in principle enough to solve the metric. However, it turns out that it
is useful to make use also of the $\theta\theta$ component of the metric equation
\be
\frac{\zeta''+\nu''}{2}+\frac{(\nu')^2}{4}+\frac{\zeta'+\nu'}{2r}=-\frac{\kappa (Gm_s)^2}{4 \pi }
\frac{e^{-(\zeta+\nu)}}{r^4}+K\frac{(\nu')^2}{8}.
\label{eq3}
\ee 
The study of the properties of these  equations for the appropriate boundary conditions 
constitutes most of the rest of this work.    

\subsection{Case II: the vector field has a non-vanishing r component}

 In the non-aligned case, $U^r\ne 0$,  the 
Lagrange multiplier $\lambda_{\rm II}$ is given by [see Eq.~(\ref{Ur})]
\be   
\lambda_{\rm II}=-\frac{\kappa (Gm_s)^2}{2\pi}\frac{e^{-(\nu+4\zeta)}}{r^4[e^{-\zeta}-(U^r)^2]^2}.
\label{lambdaII}
\ee
The components of the vector field are connected to the functions $\nu$ and $\zeta$ through
the normalization Eq.~(\ref{norm})
\be
e^\nu(U^t)^2-e^\zeta(U^r)^2=1
\label{normII}
\ee
and the $t$ component of the vector equation (\ref{vectoreq}) (given in a compact form) yields 
\be
KU^{[t;\beta]}{}_{;\beta}-\frac{\kappa (Gm_s)^2}{2\pi}\frac{e^{-(\nu+4\zeta)}}
{r^4[e^{-\zeta}-(U^r)^2]^2} U^t=0,
\label{utII}
\ee
where the $U^{[t;\beta]}{}_{;\beta}$ term involves derivatives of the four unknown functions
$\nu$, $\zeta$, $U^r$, $U^t$. 

The last two expressions can be combined with the $tt$ and $rr$
components of the metric equation (\ref{metrice}) to arrive to a closed system of 
four differential equations with four unknown functions. To this end, one has to 
calculate the relevant $\tau_{\alpha\beta}$ and $\Theta_{\alpha\beta}$ terms.
Equation (\ref{tau}) yields
\be
\tau_{tt}=\frac{\kappa Gm_s^2}{32 \pi^2 }\frac{e^{-3\zeta}}{r^4[e^{-\zeta}-(U^r)^2]},
\ee
\be
\tau_{rr}=\frac{\kappa Gm_s^2}{32 \pi^2 }\frac{e^{-(\nu+3\zeta)}}{r^4[e^{-\zeta}-(U^r)^2]^2}
\Big(1-3(U^r)^2\Big).
\ee
For $\Theta_{tt}$ and $\Theta_{rr}$ we have [see Eq.~(\ref{Theta})]
\be
\Theta_{tt}=\frac{K}{2}e^{2\nu-\zeta}[\nu'U^t+(U^t)']^2+\frac{\kappa (Gm_s)^2}{2 \pi }
\frac{(U^t)^2e^{\nu-4\zeta}}{r^4[e^{-\zeta}-(U^r)^2]^2},
\ee
\be
\Theta_{rr}=-\frac{K}{2}e^{\nu}[\nu'U^t+(U^t)']^2+\frac{\kappa (Gm_s)^2}{2 \pi }
\frac{(U^r)^2e^{-(\nu+2\zeta)}}{r^4[e^{-\zeta}-(U^r)^2]^2}.
\ee
The task in the next subsection is to study the asymptotic behavior of the physical metric
far from the source in the two cases I and II and derive the Post-Newtonian corrections
predicted by TeVeS.      

\subsection {Asymptotic behavior of the metric far from the source}

Far from the source (but not too far, so that the MOND corrections can be safely neglected),
the  metric can be taken to be asymptotically flat. Expanding the $e^\zeta$, $e^\nu$ 
to powers of $r/r_g$ (where $r_g$ is a length scale to be determined) we have
\be
e^\nu=1-r_g/r+ a_2 (r_g/r)^2+\cdots 
\label{exp1} 
\ee  and \be
e^\zeta=1+b_1 r_g/r+ b_2 (r_g/r)^2+\cdots,
\label{exp2}
\ee 
where the proportionality constant of the second term in the expansion  (\ref{exp1})
has been absorbed by $r_g$.
We now proceed to calculate the coefficients $a_i$ and $b_i$ of the metric and 
equivalent coefficients of the physical metric $\tilde{g}_{\alpha\beta}$ for the
two  cases I, II (defined in the previous Section).

\subsubsection{Case I: $U^r=0$}

If the vector field is aligned with the time direction 
one can substitute the expansions (\ref{exp1}), (\ref{exp2}) into the metric equations
(\ref{eq1}) and (\ref{eq2}),  match coefficients
of like powers if $1/r$ and solve for the 
coefficients $a_{i}$, $b_{i}$. Doing so to order of $(1/r)^3$,
the metric has the form 
\be
e^\nu=1-\frac{r_g}{r}+\frac{1}{2}\frac{r_g^2}{r^2}-\frac{1}{96}\Big[18+\frac{2\kappa}{\pi}
\Big(\frac{Gm_s}{r_g}\Big)^2-K\Big] \frac{r_g^3}{r^3}
\label{boundary1}
\ee
 and 
\begin{eqnarray}
e^\zeta=1+\frac{r_g}{r}+\frac{1}{16}\Big[6-\frac{2\kappa}{\pi}\Big(\frac{Gm_s}{r_g}\Big)^2+K\Big]
\frac{r_g^2}{r^2}+\frac{1}{96}
\Big[6-\frac{10\kappa}{\pi}\Big(\frac{Gm_s}{r_g}\Big)^2+5K\Big]\frac{r_g^3}{r^3}.
\label{boundary2}
\end{eqnarray}
In this expansion, one can see that the first
corrections introduced by TeVeS with respect to the Schwarzschild metric appear in the
$(r_g/r)^2$ term in $e^\zeta$ and in the $(r_g/r)^3$ term in $e^\nu$.

Actually, these asymptotic expansions differ from expressions given in~\cite{Bek}
(compare Eqs.~(\ref{boundary1}) and (\ref{boundary2}) of this work with  Eqs. (89)-(91) in Section V 
of~\cite{Bek}).  The reason for this difference is a sign error in the $\beta_1$ term of 
the Lagrange multiplier in Bekenstein's Eq.~(82) (see also the erratum of Ref.~\cite{Bek}). 
Because of this 
discrepancy, we need to rederive the Post-Newtonian
corrections predicted by TeVeS. The physical metric $\tilde{g}_{\alpha\beta}$ is given
by the expressions $\tilde{g}_{tt}=g_{tt}e^{2\phi}$, $\tilde{g}_{ii}=g_{ii}e^{-2\phi}$; so
we still need the asymptotic behavior of $\phi$. Integrating (\ref{phiprime}) and using
Eqs. (\ref{boundary1}) and (\ref{boundary2}), we have
\be
\phi(r)=\phi_c-\frac{\kappa Gm_s}{4\pi r}-\frac{\kappa Gm_s}{192\pi}\Big[1+\frac{\kappa}{\pi}
\Big(\frac{Gm_s}{r_g}\Big)^2-\frac{K}{2}\Big]\frac{r_g^2}{r^3}+\mathcal{O}({r^{-5}}),
\ee 
where $\phi_c$ is the cosmological value of $\phi$ at a specific epoch, which can be absorbed by
rescaling of the $t$ and $r$ coordinates: $t'=te^{\phi_c}$ and $r'=re^{-\phi_c}$. Doing so and 
dropping the primes for simplicity in the notation, the physical metric is
\be
\tilde{g}_{tt}=-1+\Big(\frac{\kappa Gm_s}{2\pi}+r_g\Big)\frac{1}{r}-\frac{1}{8}\Big(2r_g+
\frac{\kappa m_s}{\pi}\Big)^2\frac{1}{r^2}+\frac{1}{192}\Big(2r_g+\frac{\kappa Gm_s}{\pi}\Big)
\Big[4\Big(\frac{\kappa Gm_s}
{\pi r_g}+2\Big)^2+2+\frac{2\kappa}{\pi}\Big(\frac{Gm_s}{r_g}\Big)^2-K\Big]\frac{r_g^2}{r^3}+\cdots
\ee
and
\be
\tilde{g}_{rr}=1+\Big(\frac{\kappa Gm_s}{2\pi}+r_g\Big)\frac{1}{r}+\frac{1}{16}\Big[2\Big(\frac
{\kappa Gm_s}{\pi r_g}+2\Big)^2-2-\frac{2\kappa}{\pi}\Big(\frac{Gm_s}{r_g}\Big)^2+K \Big]
\frac{r_g^2}{r^2}+\cdots
\ee
Identifying the $1/r$ term of the $tt$ component with $2G_N m/r$ (where $G_N$ is 
Newton's constant), the physical metric can be brought into the form
\be
\tilde{g}_{tt}=-1+2G_N m/r-2G_N^2m^2/r^2+\cdots
\ee 
and
\be
\tilde{g}_{rr}=1+2G_N m/r+\cdots
\ee
which is identical to GR up to order of post-Newtonian corrections. This means that
one has to go to higher order terms in TeVeS to obtain the corrections to GR. No 
constrains can be set to the parameters of TeVeS $\kappa$ and $K$ from 
measurements of the standard Post-Newtonian coefficients, if the radial component of the
vector field vanishes.

At this point, one more comment is in order. By inspection of Eqs.~(\ref{boundary1}), 
(\ref{boundary2}) one notices that the 
quantity $\frac{\kappa}{\pi} (Gm_s/r_g)^2-K/2$ (times some factor) appears in all the
corrections introduced by TeVeS with respect to equivalent the general relativistic solution. 
This quantity will also appear in the analytic solutions derived in Sec.~V. 

\subsubsection{Case II: $U^r\ne 0$}
 
In the non-aligned case one needs to consider the asymptotic expansion of the vector field
components. For $r_g/r\ll 1$ the vector field relaxes to its cosmological value, i.e.,
$U^t\to 1$ and $U^r\to 0$ (since there is no preferred spatial direction).
So, expanding to powers of $r_g/r$, we have
\be 
U^t=1+c_1r_g/r+c_2(r_g/r)^2+\cdots
\label{boundary3}
\ee
and   
\be
U^r=d_1r_g/r+d_2(r_g/r)^2+\cdots.
\label{boundary4}
\ee
From this point on the method we follow to calculate the Post-Newtonian corrections
is similar to this of the previous subsection. Substituting these expansions and 
Eqs.~(\ref{exp1}), (\ref{exp2}) into Eqs.~(\ref{normII}), 
(\ref{utII}) and the $tt$ and $rr$ components of the metric equations 
(\ref{metrice}) and  matching coefficients
of like powers if $1/r$, we derive the coefficients $a_i$, $b_i$, $c_i$ and $d_i$.
This analysis is carried out down to the order necessary to calculate the 
post-Newtonian coefficients and gives for $K\ll 1$, $\kappa\ll 1$
\be a_2=\frac{1}{2}+\frac{\kappa(Gm_s)^2}{4\pi r_g^2}+\frac{K}{8}, \ee 
\be b_1=1, \ee \be
b_2=\frac{3}{8}-\frac{3}{8}\frac{\kappa (Gm_s)^2}{\pi r_g^2}-\frac{K}{16},     
\ee
\be
c_1=\frac{1}{2}, \ee \be
c_2=\frac{1}{16}\Big(5+\frac{4\kappa (Gm_s)^2}{K\pi r_g^2}\pm \sqrt{\frac{8\kappa(Gm_s)^2}
{K\pi r_g^2}+5}\Big) 
\ee and \be
d_1=\frac{1}{4}\Big(1\pm \sqrt{\frac{8\kappa(Gm_s)^2}{K\pi r_g^2}+5}\Big).
\label{d1}
\ee
The $\pm$ sign in the last two coefficients comes from the fact that the normalization
expression (\ref{normII}) contains squares of the vector components. The $+$ sign
corresponds to $U^r>0$ and vice versa. 

The asymptotic behavior of the scalar field is found after expanding
and integrating Eq.~(\ref{phiprime})
\be
\phi(r)=\phi_c-\frac{\kappa Gm_s}{4\pi r}+\mathcal{O}({r^{-3}}).  
\ee
The physical metric is given by Eq.~(\ref{physmetric}) and a rescaling of the $r$, $t$ coordinates
by $t'=te^{\phi_c}$ and $r'=re^{-\phi_c}$ is needed so that it can asymptote to the
Minkowskian form. Notice, however, that $\phi_c$ is not absorbed by this rescaling unlike the
$U^r=0$ case, because of the more complicated connection of the physical metric with the
fields of TeVeS in this case. Assuming again $K\ll 1$, $\kappa\ll 1$ and furthermore that $\phi_c \ll 1$,
we have for the standard  $\beta$, $\gamma$ Post-Newtonian coefficients, as predicted by TeVeS for
the case that $U^r\ne 0$
\be
\beta=1+\frac{\kappa}{8\pi}+\frac{K}{4}+\phi_c\Big(3+\frac{\kappa}{\pi K}\pm 
\sqrt{\frac{2\kappa}{\pi K}+5}\Big)
\label{betaII}
\ee
and
\be
\gamma=1.
\ee 
Here again the $\pm$ sign in the expression for $\beta$ is determined by the sign of $U^r$
[see Eq.~(\ref{d1})]. 

While the $\gamma$ coefficient coincides with the GR prediction, the $\beta$ differs from
unity. The best determination of $\beta$ comes from lunar laser ranging tests (see for
example Ref.~\cite{Anderson}) which in combination with the value for $\gamma$ measured by the
Cassini experiment \cite{Bertotti} yields $\beta-1\simless 10^{-4}$ \cite{Williams}.     
How does this result compare with the prediction of Eq.~(\ref{betaII})? 
It is important to note that the term multiplied with the (positive) $\phi_c$ in Eq.~(\ref{betaII})
is positive for any value of $\kappa/K>0$ and choice of the $\pm$ sign, so one has the inequality
$\beta-1\ge \kappa/(8\pi)+K/4$. For the choice of the function $F$ made in Ref.~\cite{Bek},
$\kappa$ is constrained to be $\simeq 0.03$ which results in $\beta-1\simmore 2.5\cdot 10^{-3}$
(taking the $K$ term much smaller). This is in conflict with observations. 

Summarizing, in this section we have shown that if the vector field is aligned with the time
direction, the standard Post-Newtonian coefficients derived by TeVeS are identical to these of
GR, while if $U^r\ne 0$, the PPN correction for the $\beta$ coefficient is in 
 conflict with best determinations of $\beta$. This means that either one 
has to assume that $U^r=0$, or  a different choice of the function $F$ than that
of Ref.~\cite{Bek} has to be made so that TeVeS is in accordance with solar system phenomenology.         

\section{Analytic solutions when $U^r$ vanishes}

Until now, we have kept the study of spherical symmetric spacetimes in TeVeS quite general.
From this point on, we focus on the branch of solutions for which $U^r=0$, i.e.,
the vector field is aligned to the time direction. As it turns out, exact analytic 
solutions are possible in this case.

\subsection{Solutions in the $K\to 0$ limit}

The system of Eqs.~(\ref{eq1}), (\ref{eq2}) and (\ref{eq3}) is rather complicated. 
Here, we first consider some special cases and then use the
intuition we gain to derive the general solution.
In the simplest case where both $\kappa=0$ and $K=0$, the metric equations
in TeVeS coincide with these in GR and their write hand side is zero
(i.e. no source terms appear). In this limit the integration of Eqs.~(\ref{eq1}), 
(\ref{eq2}) is straightforward leading to the familiar GR solution
\be 
e^\nu=\Big(\frac{1-r_g/4r}{1+r_g/4r}\Big)^2,
\label{grsol1}
\ee
\be
e^\zeta=(1+r_g/4r)^4,
\label{grsol2}
\ee
where the boundary conditions (\ref{boundary1}), (\ref{boundary2}) have been used. 
In this  case, one can show [see Eq.~(\ref{phiprime})] that $\phi$ is constant at its 
cosmological value $\phi_{c}$ and that $g_{\alpha\beta}$  coincides with 
that predicted by GR. The physical metric is given by
 $\tilde{g}_{tt}=g_{tt}e^{2\phi_c}$, $\tilde{g}_{ii}=g_{ii}e^{-2\phi_c}$.
 The factors $e^{\pm 2\phi_c}$ can be absorbed by an appropriate rescaling of the $t$ and $r$
coordinates, resulting in a physical metric is equivalent to that of GR.

As a next step toward the most general solution, we take the limit $K\to 0$ but allow
$\kappa$ to be arbitrary. In this limit we essentially decouple the theory from the
vector field and the metric equations become
\be   
\zeta''+\frac{(\zeta')^2}{4}+\frac{2\zeta'}{r}=-\frac{\kappa (Gm_s)^2}{4 \pi }\frac{e^{-(\zeta+\nu)}}
{r^4},
\label{eq1s}
\ee
\be 
\frac{(\zeta')^2}{4}+\frac{\zeta'\nu'}{2}+\frac{\zeta'+\nu'}{r}=\frac{\kappa (Gm_s)^2}{4 \pi }
\frac{e^{-(\zeta+\nu)}}{r^4}
\label{eq2s}
\ee
and
\be
\frac{\zeta''+\nu''}{2}+\frac{(\nu')^2}{4}+\frac{\zeta'+\nu'}{2r}=-\frac{\kappa (Gm_s)^2}{4 \pi }
\frac{e^{-(\zeta+\nu)}}{r^4}.
\label{eq3s}
\ee
It turns out that that Eqs.~(\ref{eq1s}), (\ref{eq2s}), (\ref{eq3s})  are equivalent to 
 spherical symmetric spacetimes in metric-massless scalar theories of gravity. 
The exact solution was originally
written down by Buchdahl in Ref.~\cite{Buchdahl} (see also Ref.~\cite{Mannheim}). Here, we will briefly
repeat the derivation.

From the addition of Eqs.~(\ref{eq2s}) and (\ref{eq3s}) we find
\be
2(\nu''+\zeta'')+(\nu'+\zeta')^2+6\frac{\nu'+\zeta'}{r}=0,
\ee
which can be integrated once to give
\be
\nu'+\zeta'=\frac{4r_c^2}{r(r^2-r_c^2)}.
\label{nuzetaprime}
\ee
Where we have introduced the integration constant $r_c^2$. 
This constant can be evaluated by expanding Eq.~({\ref{nuzetaprime})
to powers of $1/r$ and comparing with the expansions (\ref{boundary1})
and (\ref{boundary2}). After some algebra we find
\be
r_c=\frac{r_g}{4}\sqrt{1+\frac{\kappa}{\pi}\Big(\frac{Gm_s}{r_g}\Big)^2}.
\label{rc}
\ee
Equation ({\ref{nuzetaprime}) can be integrated again to yield
\be
\nu+\zeta=2\ln\Big(\frac{r^2-r_c^2}{r^2}\Big),
\label{nuzeta}
\ee
where the second integration constant has been set to unity so that the 
asymptotic form of the solution is a flat spacetime (i.e.
$e^{\nu+\zeta}\to 1$ for $r\to \infty$).

One verifies that after setting
\be
\zeta'=\frac{4r_c^2}{r(r^2-r_c^2)}-\frac{r_g}{r^2-r_c^2}
\ee
and using  Eq. (\ref{nuzeta}) to derive $\nu'$, the metric equations are all
satisfied. After integrating for $\nu$ and $\zeta$, one has the exact solution
for the metric components
\be
e^\nu=\Big(\frac{r-r_c}{r+r_c}\Big)^{r_g/2r_c}
\label{ssol1}
\ee
and
\be
e^\zeta=\frac{(r^2-r_c^2)^2}{r^4}\Big(\frac{r-r_c}{r+r_c}\Big)^{-r_g/2r_c},
\label{ssol2}
\ee
where $r_c$ is given by Eq.~(\ref{rc}). It is straightforward to check that
in the limit where $\kappa m_s\to 0$, one derives the well known
general relativistic expressions. 

Having solved for the metric components, one can integrate Eq.~(\ref{phiprime}) to 
derive the $r$ dependence of the scalar field and then the physical metric through 
Eq.~(\ref{physmetric}). However, the results derived in this Section are of limited
generality since they correspond to the $K=0$ case, where the effect of the
vector field to the metric equations is ignored. The generalization of the solutions
to the case where $K\ne 0$ is the task of the next Section.

\subsection{Spherically symmetric, vacuum solution for the metric for
arbitrary $\kappa$, $K$} 

We turn to the general case where both $\kappa$ and $K$ are non-zero.
While at first sight the metric equations look quite complicated in this
case, it turns out that  one can repeat the procedure of the
previous Section to derive more general spherical symmetric, vacuum solutions
for the metric which are identical to (\ref{ssol1}), (\ref{ssol2}) provided that one
makes the substitution 
\be
\frac{\kappa}{\pi}\Big(\frac{Gm_s}{r_g}\Big)^2 \to \frac{\kappa}{\pi}\Big(\frac{Gm_s}{r_g}\Big)^2
-\frac{K}{2} 
\ee
in the definition of $r_c$ [Eq. (\ref{rc})], i.e.,
\be
r_c=\frac{r_g}{4}\sqrt{1+\frac{\kappa}{\pi}\Big(\frac{Gm_s}{r_g}\Big)^2-\frac{K}{2}}.
\label{rc2}        
\ee

In the  $\kappa$, $K$ parameter plane the line defined by 
\be
K=\frac{2\kappa}{\pi}\Big(\frac{Gm_s}{r_g}\Big)^2
\label{expr}
\ee 
has a special significance. 
With Eq.~(\ref{expr}) $r_c=r_g$ and the metric is identical to
the general relativistic one \footnote{On the other hand, one should keep in mind that
the physical metric $\tilde{g}_{\alpha\beta}$ differs by a factor $e^{\pm 2\phi}$
from $g_{\alpha \beta}$}. This is a result of the fact that the energy density
contributed by the scalar is exactly canceled out by the \emph{negative} energy density
of the $\Theta_{tt}$ in the right hand side of the $tt$ component of Eq.~(\ref{metrice}),
i.e. $8\pi G\tau_{tt}+\Theta_{tt}=0$. Actually, when
\be
K>\frac{2\kappa}{\pi}\Big(\frac{Gm_s}{r_g}\Big)^2
\ee
the total energy density of vacuum contributed by the fields is negative in the
whole space time.  This can have important consequences for the theory
since it may lead to instability of the vacuum from the quantum point of
view.

The behavior of the scalar field  can be
followed by integrating Eq. (\ref{phiprime}) and the use of Eq. 
(\ref{nuzeta}) 
\be 
\phi(r)=\phi_c+\frac{\kappa G m_s}{8\pi r_c}\ln \Big(\frac{r-r_c}{r+r_c}\Big),
\label{phi}
\ee 
where $\phi_c$ stands for the cosmological value of the scalar field at a specific
epoch. Just as in metric-scalar theories, one can see that, unless $\kappa m_s=0$ 
the scalar field diverges logarithmically at $r=r_c$ and that there is always a radius 
$r_1>r_c$ where $\phi(r_1)=0$ and becomes negative further in.
We will return to this point in the next Section
where we discuss how black holes look like in TeVeS, and how much they
differ from the ones predicted by GR.

The components of physical metric is related to $\nu$, $\zeta$ and $\phi$ through the expressions 
(\ref{physmetric}), (\ref{ssol1}), and (\ref{ssol2}) that yield
\be
\tilde{g}_{tt}=-\Big(\frac{r-r_c}{r+r_c}\Big)^a
\label{phystt}
\ee
and
\be
\tilde{g}_{rr}=\frac{(r^2-r_c^2)^2}{r^4}\Big(\frac{r-r_c}{r+r_c}\Big)^{-a},
\label{physrr}
\ee
where $a\equiv \frac{r_g}{2r_c}+\frac{\kappa Gm_s}{4\pi r_c}$.

The expressions (\ref{phystt}) and ({\ref{physrr}) describe spherically symmetric, vacuum
spacetimes, i.e., they describe the spacetime down to the surface of a star.   
The two dimensionless parameters $\kappa$ and
$K$ of the theory provide the parameter space that is to be explored.
In addition to these, we have kept the scalar mass $m_s$ and the gravitational
radius $r_g$ as free parameters so that the derived results to be quite
general and applicable to both the case of a black hole and the exterior of a star.
In appendix D of Ref.~\cite{Bek}  a detailed description of  the procedure to calculate 
$m_s$ and $r_g$ in terms of its
gravitational mass $m_g$ of the star is given. Unfortunately, this method is not applicable to 
the case of a black hole and a different approach is needed to determine $m_s$
and $r_g$.

\section{How do black holes look like in TeVeS?}

 The characteristic radius of the physical metric described by Eqs.~(\ref{phystt})
and (\ref{physrr}) is $r_c$. At $r_c$ the $tt$ component of the metric vanishes and
the question that rises is whether and under which conditions $r_c$ is the location 
of the horizon of a black hole. A first step toward answering this question is 
to calculate the surface area at this radius, which turns out to be proportional to 
$\tilde{g}_{rr}(r_c)$. A black hole must have a finite 
surface area at $r_c$  which [in view of Eq. (\ref{physrr})] constrains $a$ to be $\le 2$.   

A second constraint on $a$ (Bekenstein private communication) comes from the 
demand that there is no essential singularity at $r_c$. For our solution, the Ricci scalar
$R$ is
\be
R=\frac{2(a^2-4)r_c^2r^4(r-r_c)^{a-4}}{(r+r_c)^{a+4}}.
\ee  
From this expression one can see that the Ricci scalar is finite when
$a=2$ or $a>4$. Considered together, the two constraints (i.e. of finite surface area and
Ricci scalar at $r_c$) imply that only the value $a=2$ describes a black hole. 
Using the definition of $a$ we have that for $a=2$
\be
r_c=\frac{r_g}{4}+\frac{\kappa Gm_s}{8\pi}
\label{rcbh}
\ee   
and the physical metric has the form
\be
\tilde{g}_{tt}=-\Big(\frac{r-r_c}{r+r_c}\Big)^2
\label{physttbh}
\ee
and
\be
\tilde{g}_{rr}=\Big(1+\frac{r_c}{r}\Big)^4.
\label{physrrbh}
\ee
This is exactly the GR solution after setting $r_c=G_Nm/2$. 
So, the physical metric in TeVeS
is identical to this of GR for a non-rotating black hole.

Furthermore, one can use the definition of $r_c$ [see Eq.~(\ref{rc2})]
in Eq.~(\ref{rcbh}) to solve for $m_s$ and finds
\be
\frac{Gm_s}{r_g}=\frac{1+\sqrt{1+\Big(2-\frac{\kappa}{2\pi}\Big)\frac{\pi K}{\kappa}}}
{2-\frac{\kappa}{2\pi}}.
\ee    
We have already shown in the previous Section that when $m_s\ne 0$, there is 
always a region close to $r_c$ where the scalar field turns negative. 
Bekenstein in ref.~\cite{Bek}, on the other hand, has shown that TeVeS becomes acausal
(i.e. it suffers from superluminal propagation of metric, vector and scalar
field disturbances) when $\phi<0$. As a result, the theory appears to behave in an unphysical 
way  in the vicinity of our  black hole solution. On the other hand, our solutions have been 
derived under the assumption that $U^r=0$. Perhaps the causality problem can be overcome by 
allowing for $U^r\ne 0$.

\section{Conclusions}

Bekenstein's recent relativistic gravitational theory 
(TeVeS) that leads to MOND in the relevant limit has been proposed
as a modification to GR. TeVeS has several attractive features, for
example it predicts the right amount of gravitational lensing
when only the observed mass is used and provides a covariant formulation to 
construct cosmological models.

The free parameters in TeVeS can be constrained by the large extra-galactic phenomenology.
In this work, instead, we  have looked at TeVeS in the strong gravity limit. Two branches of 
solutions are identified: the first is characterized by the vector 
field being aligned with the time direction while in the (not previously explored) second branch
the vector filed has a non-vanishing radial component.  We have shown that the $\beta$ and $\gamma$ 
PPN coefficients in TeVeS are identical to these of GR in the first branch of solutions while the 
$\beta$ PPN coefficient differs in the two theories in the second. Despite the fact that the
results derived here are essentially independent of the exact choice of the free function $F$ of the theory,
 our final conclusions do depend on it, since the choice of $F$ influences the way in which observations put
 constraints on the parameters of the theory. For the second branch of solutions and for the choice 
of the free function $F$ made in Ref.~\cite{Bek}, TeVeS predicts $\beta$ that is in conflict with 
recent observational determinations of it. 

For the first branch of solutions, we derive analytic 
expression for the physical metric. These solutions are an extension of these that describe
spherical symmetric spacetimes in tensor-massless scalar theories and depend on the values of
the two dimensionless parameters $\kappa$, $K$ of TeVeS and the ratio $Gm_s/r_g$.
One of the findings of this work is that the energy density contributed by the vector field 
is negative, and when $K>\frac{2\kappa}{\pi}(Gm_s/r_g)^2$ the total
energy density of vacuum also becomes negative, possibly turning it unstable from
the quantum point of view.

In the case of a black hole, our solutions for the metric are identical to the Schwarzschild 
solution in GR. On the other hand, these solutions are shown to be acausal in the vicinity
of the black hole.  Possibly, the issues of the negative energy density contributed by the
vector field, and  of  causality close to a black hole do not appear in the case where
$U^r\ne 0$. In this case, however, a different choice of the free function $F$ will be needed
so that TeVeS is not in conflict with solar system phenomenology.

\begin{acknowledgements}

I thank Demos Kazanas for bringing exact solutions of metric-scalar theories
of gravity to my attention and Jacob Bekenstein and Henk Spruit for valuable suggestions and comments 
on a previous version of the manuscript. This work was supported by the EU FP5 Research Training Network 
``Gamma Ray Bursts: An Enigma and a Tool.''

\end{acknowledgements}

\end{document}